# A method for determining the V magnitude of asteroids from CCD images

## Roger Dymock & Richard Miles

We describe a method of determining the V magnitude of an asteroid using differential photometry, with the magnitudes of comparison stars derived from *Carlsberg Meridian Catalogue 14* (CMC14) data. The availability of a large number of suitable CMC14 stars enables a reasonably accurate magnitude (±0.05 mag) to be found without having to resort to more complicated absolute or all-sky photometry. An improvement in accuracy to ±0.03 mag is possible if an ensemble of several CMC14 stars is used. This method is expected to be less accurate for stars located within ±10° of the galactic equator owing to excessive interstellar reddening and stellar crowding.

## The problem

Differential photometry is fairly straightforward in that all the required data can be obtained from images of the target object. Changing atmospheric conditions and variations in dimming due to altitude are likely to affect all stars in the image equally and can thus be ignored. However, deriving asteroid magnitudes in this way is problematic in that comparison stars with known accurate magnitudes are few and far between, and it is unusual to find many, if any, such stars on a typical random CCD image. So to obtain an accurate measure of the magnitude of the target asteroid, one must ordinarily resort to 'absolute' or 'all-sky' photometry. This approach is much more complicated than differential photometry: the sky must be adequately clear (sometimes referred to as 'photometric'); standard stars need to be imaged (usually some distance from the target asteroid); at least one filter must be used; and extinction values, nightly zeropoints, airmass corrections and the like must all be taken into account.

## Proposed solution

### Overview

From a single image, or preferably a stack of several images, the Johnson V magnitude (centred on ~545nm) of an asteroid may be obtained, accurate to about ±0.05 mag. All that is required is access to the *Carlsberg Meridian Catalogue 14* (CMC14) from which data the magnitudes of the comparison stars can be calculated. We set out here a method for deriving such V magnitudes of comparison stars in the range 10<V<15. It is possible that a more automated approach could be devised by someone with sufficient computer skills (see Postscript for significant developments in this respect since this paper was first written).

### The CMC14 catalogue

The Carlsberg Meridian Telescope (formerly the Carlsberg Automatic Meridian Circle) is dedicated to carrying out high-precision optical astrometry.[1] It underwent a major upgrade in 1999 March with the installation of a 2k×2k pixel CCD camera together with a Sloan Digital Sky Survey (SDSS) r' filter, the system being operated in drift-scan mode.[2] With the new system, the r' magnitude limit is 17 and the positional accuracy is in the range 35–100 milliarcsec. The resulting survey is designed to provide an astrometric and photometric catalogue in the declination range −30 to +50°. The CMC14 catalogue is the result of all the observations made between 1999 March and 2005 October in this declination band, except for a gap between 05h30m and 10h30m for declinations south of −15°. It contains 95,858,475 stars in the Sloan r' magnitude range 9 to 17. The Sloan r' band is in the red part of the spectrum centered at ~623nm and having a bandwidth at half-maximum of 137nm.

Importantly, a VizieR query[3] of the CMC14 catalogue returns the r' magnitude data and also the Two-Micron All-Sky Survey (2MASS) data comprising the J-band (1.25 μm), H-band (1.65 μm), and $K_s$-band (2.17 μm) magnitudes.[4] To be able to convert the known r' magnitude of a star into a standard V magnitude, we must know the colour of the star. The difference in the J and K magnitudes provides such a measure of star colour.

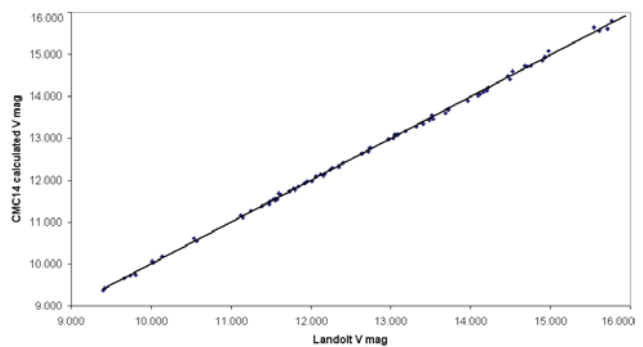

**Figure 1.** Landolt vs CMC14 derived V magnitudes.

## Theory

In 2006, John Greaves analysed data for 696 stars (9.9<V<14.8) from the LONEOS photometric database produced by Brian Skiff.[5] He merged these data with r' magnitude data from the earlier Carlsberg CMC12 catalogue and with 2MASS data (0.00<(J–K)<1.00) and found that the relationship

$$V = 0.641*(J-K) + r' \qquad [1]$$

predicted the V magnitude with fair accuracy (standard deviation of 0.038 mag) when comparing V(Loneos) – V(calculated) using the above linear relationship.[6]

It was found that a limit has to be placed on the allowable values of (J–K) since the above relationship begins to break down for very red stars. From further studies, John Greaves and Richard Miles concluded that the choice of comparison stars should be limited to those with a value of (J–K) between 0.3 and 0.7. However, where there is only one comparison star in the image the allowable J–K range might have to be relaxed to between 0.2 and 0.8.

In a separate exercise, the V magnitudes of one hundred Landolt stars, the accepted standard for visible photometric calibration, were calculated using Equation [1] above and a graph of those derived values plotted against the Landolt magnitudes.[7] Subsequent analysis of the data by the present authors led to a 'best-fit' modification of the above formula to:

$$V = 0.6278*(J-K) + 0.9947*r' \qquad [2]$$

A graph of V magnitude derived from the CMC14 catalogue versus Landolt magnitude is shown in Figure 1. The mean error of the CMC14-derived V magnitudes was calculated to be 0.043 for stars brighter than V=14. Plotting residuals as shown in Figure 2 clearly illustrates the departure of the calculated magnitude from the Landolt values based on Equation [2] above. Throughout the magnitude range, 9< r'<16, the average correlation shows no systematic trend away from linearity, only a slightly increased scatter at V >15.

Also shown plotted in Figure 2 are the differences between CMC-derived V magnitudes and those calculated from *Tycho-2* data using *GUIDE 8.0* software respectively.[8,9] The *Tycho* catalogue created as part of the *Hipparcos* space mission comprises about 2.5 million stars, and has been used by many as a source of magnitude data. However, it can be seen from Figure 2 that there is significant scatter for the *Tycho* data at magnitudes fainter than about 10.5, showing that this commonly used catalogue is unsuitable for use as a source of accurate photometry for stars fainter than this limit. Un-like *Tycho* the CMC14 catalogue provides a good reference source for field calibration down to 15th magnitude.

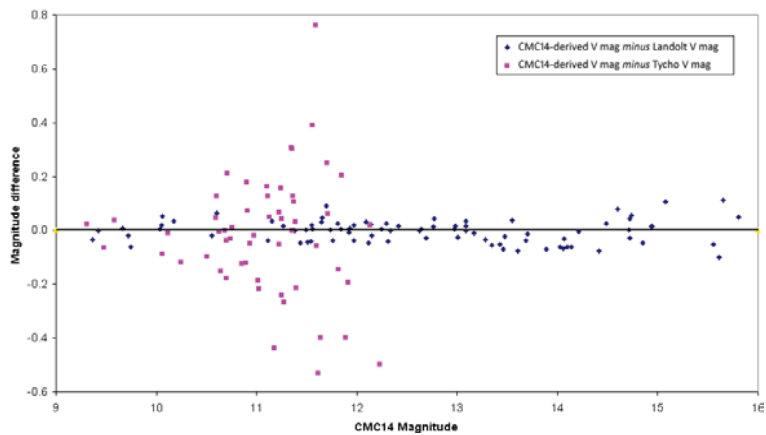

**Figure 2.** Residual plot of Landolt vs CMC14-derived magnitudes and Tycho V magnitudes.

## Practical results

The equipment used by the authors is listed in Table 1.

### Example 1: Determining the V magnitude of asteroid (1467) Mashona

Imaging, calibration, and magnitude calculation described in this example were carried out by RD as detailed below. Asteroid *(1467) Mashona,* imaged on 2007 October 17, is used as the example.

#### Guidelines

Here are some guidelines to follow if photometry of asteroids is planned:

− Choose asteroids higher than 25–30° altitude.
− Do not attempt to image objects that are too faint to ensure a sufficiently high signal-to-noise ratio (SNR), ideally >20: in RD's case using a 0.25m telescope this means working on objects brighter than V=15. The SNR can be improved by stacking multiple images.
− Choose an exposure time which avoids excessive trailing (say by more than 2 or 3 pixels) due to the motion of the asteroid.
− Check the stacked images to ensure that the asteroid and comparison stars do not become saturated. Where necessary stack in 'Average' mode to ensure saturation is avoided.
− Check for suitable comparison stars in the field of view (FOV): in RD's case at least two stars brighter than V=14. A planetarium program such as

**Table 1. Instrumentation used**

| Item | Roger Dymock | Richard Miles |
|---|---|---|
| Telescope | Orion Optics 25cm f/6.4 Newtonian on a German equatorial mount | Celestron 28cm f/10 Schmidt–Cassegrain and Takahashi 6cm f/5.9 refractor, both on the same German equatorial mount |
| CCD camera | Starlight Xpress MX516 | Starlight Xpress SXV-H9 |
| Filter | Johnson V | Johnson V, Cousins I and unfiltered |

**Table 2.** Calculation of V magnitude for asteroid (1467) Mashona

| Col.: | A | B | C | D | E | F | G | H | I | J | K | L |
|---|---|---|---|---|---|---|---|---|---|---|---|---|
| Row | | | | | | | | | | | | |
| 1 | | AIP4WIN data | | | | CMC14 data | | | | | | |
| 2 | | v | (v–V) | Calc | GSC No. | r' | J | K | J–K | V | Error | Used |
| 3 | C1 | −6.47 | −19.67 | | 1726-1112 | 12.72 | 9.81 | 8.95 | 0.86 | 13.20 | 0.02 | No* |
| 4 | C2 | −7.34 | −19.80 | | 1726-1088 | 12.31 | 11.45 | 11.11 | 0.34 | 12.46 | 0.01 | Yes |
| 5 | C3 | −6.99 | −19.76 | | 1726-1073 | 12.48 | 11.1 | 10.54 | 0.56 | 12.77 | 0.01 | Yes |
| 6 | C4 | −6.41 | −19.77 | | 1726-1377 | 13.16 | 11.94 | 11.5 | 0.44 | 13.36 | 0.02 | Yes |
| 7 | Mean | | −19.78 | | | | | | | | | |
| 8 | Ast | −6.57 | | | | | | | | | 0.02 | |
| 9 | Asteroid V mag | | | 13.21 | | | | | | | | |
| 10 | Error – Imaging ± | | | 0.03 | | | | | | | | |
| 11 | Error – Cat ± | | | 0.03 | | | | | | | | |
| 12 | Error – Total ± | | | 0.04 | | | | | | | | |

* Comparison star C1 was not used in the calculation because it was found to possess a J–K colour index outside the acceptable 0.30–0.70 range described earlier.

*MegaStar*[10] or *GUIDE*[9] is useful for doing this and to avoid inadvertently using known variable stars as comparison stars.

– Access the CMC14 catalogue to verify that the chosen comparison stars are of a satisfactory colour, i.e. have a (J–K) value between 0.30–0.70, or 0.20–0.80 if there is only one comparison star in the FOV.

If these guidelines are not followed much time can be wasted 'chasing' unsuitable targets and determining magnitudes which may have considerable uncertainty.

### Imaging

Twenty images of the asteroid were obtained using an exposure time of 30 seconds. To avoid trailing when stacking images, the maximum time interval from start of the first image to the end of last image should not exceed a value defined using a formula proposed by Stephen Laurie, *i.e.* the maximum interval in minutes equals the full width at half maximum (FWHM) of a star image in arcseconds divided by the rate of motion of the asteroid in arcseconds per minute.[11]

For *(1467) Mashona* the motion amounted to 0.039 arcsec/min as obtained from the Minor Planet Center's Ephemeris Service,[12] and the FWHM was 4 arcsec giving a total duration of 4/0.039 = 102 min, therefore trailing would not pose a problem using a sequence of 5 images obtained over a 5-minute interval.

Calibration frames consisting of 5 dark frames, 5 flat-fields, and 5 flat-darks (dark frames having the same exposure as the flat-fields) were taken for each imaging session.

### Image processing

Master dark frames and master flat-fields were generated using the software *Astronomical Image Processing for Windows* (*AIP4WIN*).[13] Calibration frames were median-combined and saved.

Images containing the asteroid were calibrated and stacked, typically 3–5 images in each stack, to improve SNR, using the *Astrometrica* software.[14] The stacked image was measured using *AIP4WIN*. A whole series of images can be processed in one batch run using the 'Multiple Image Photometry Tool' facility. The first image in a series is used to set up the analysis which can then be automatically applied to all images in that series.

The task now is to utilise data from the CMC14 catalogue for each of the comparison stars to derive its V magnitude and, together with the *AIP4WIN* data, calculate the V magnitude of the asteroid.

### Accessing the CMC14 catalogue using Aladin

A convenient tool for carrying out the analysis is the online software, *Aladin Sky Atlas*.[15] Using this facility it is possible to display and align one's own image, a Digital Sky Survey (DSS) image and the CMC14 catalogue, in chart form as shown in Figure 3.[16]

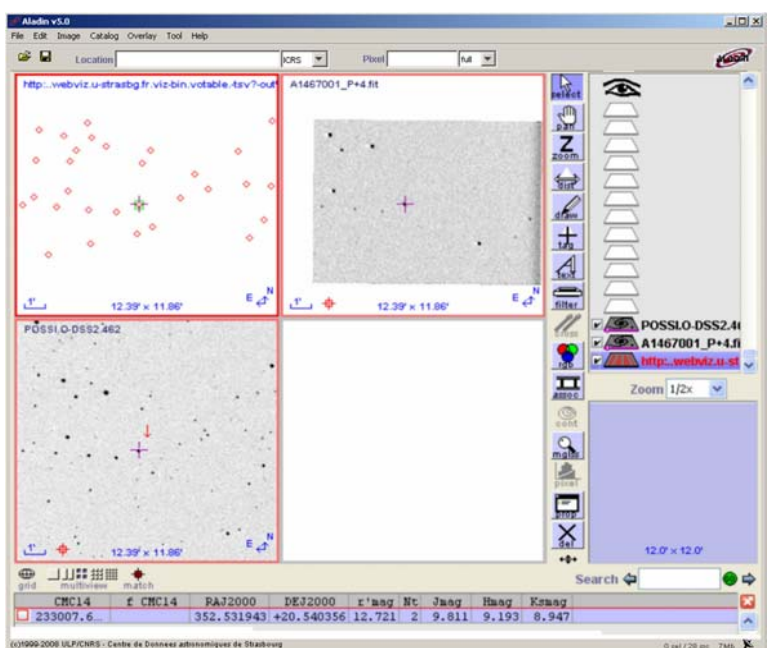

**Figure 3.** CMC14 chart overlaid on own image (left) and DSS image (right).

Performing an astrometric calibration will align the CMC14 chart with both images. Clicking on the relevant star displays the CMC14 data for that star as shown near the bottom of Figure 3. The data for each star can then be copied to, for example, an *Excel* spreadsheet.

### Magnitude calculation

The magnitude of the asteroid was calculated with the aid of a spreadsheet using the relevant CMC14 catalogue values and Equation [2] as set out below. The results are shown in Table 2.

– The values r', J and K for each star obtained from CMC14 are given in columns F–H.
– (J–K) is calculated in column I.
– The V magnitude, in column J, is calculated from the formula V = 0.6278*(J–K)+0.9947*r'
– The instrumental magnitudes, v, (from *AIP4WIN*) for each of the comparison stars, C1 to C4, and for the asteroid are given in column B.
– (v–V) for each comparison star is calculated in column C, rows 3–6.
– The mean of (v–V) for all four comparison stars is calculated in cell C7.
– The V magnitude of the asteroid is calculated in cell D9 by subtracting the mean (v–V) from the instrumental magnitude, v, of the asteroid.

### Errors

The errors are calculated as described below.

#### 1. Catalogues

Calculating the uncertainty in the results is an important part of the exercise as we are seeking to demonstrate that this methodology is significantly better than other available options (*Tycho-2*, *UCAC-2*, *USNO-A2.0*, etc.)[17,18] Many variable star observers have used the *Tycho* catalogue as a source of comparison star data. The All-Sky Survey (TASS), for example, is calibrated against *Tycho-2* stars.[19] *GUIDE* states that 'In most cases the precision provided by *Tycho* is much greater than all earlier catalogues. About the only case in which *Tycho* data would be ignored is if *Hipparcos* data is available instead'.[20] Although *Hipparcos* is indeed a very accurate catalogue for V photometry, it contains only 118,209 stars, most of which are brighter than V=9 and so are effectively too bright and too sparsely distributed to be of use as photometric reference stars in most cases.

One problem with using *Tycho* data even for the brighter stars is that there are not likely to be many such stars on any particular image. For example, a 12×8 arcmin image typically contains one *Tycho* star. A larger FOV will contain more but at high galactic latitudes, where stellar densities are much lower than the average, it is common not to have any *Tycho* star in the FOV of a CCD image. By comparison, CMC14 contains almost *40 times* as many stars as *Tycho*: hence its suitability as a potential source of comparison stars. Table 3 lists the uncertainties quoted for the magnitude of individual stars in CMC14 depending on their brightness.

**Table 3. CMC14 photometric uncertainty**

| r' | Δr' (mag) |
|---|---|
| <13 | 0.025 |
| 14 | 0.035 |
| 15 | 0.070 |
| 16 | 0.170 |

It can be seen that fainter than r'=15, the error in r' becomes significant. It is therefore best if we only use r' mags between 9 and say 14.5. In an area relatively devoid of comparison stars (say if only one is present), it may be necessary to resort to stars fainter than r'=14.5. For example, the average of a group or ensemble of say four or five r'=15.5 stars may be as good as a single star brighter than r'=14.5. Typically, the magnitude error for an ensemble of reference stars is given by the catalogue error divided by (square root of no. of stars used *minus* 1). Since Equation [2] also includes the (J–K) term, errors in the *2MASS* catalogue will also contribute towards the uncertainty in the derived V magnitude. Note that for greater accuracy, the colour range of comparison stars is restricted to those with 0.3 < (J–K) < 0.7.

#### 2. Signal to noise ratio

The errors (standard deviation or sigma) as reported by *AIP4WIN* represent only some of the sources of uncertainty in that they are based on the photon counts for the star and sky background and do not take into account many other factors affecting accuracy such as variations in sky transparency, reference catalogue errors, accuracy of flat-fields, *etc*. Using the method described in the AAVSO *CCD Observing Manual Section 4.6*,[21] the instrumental magnitudes of stars over a range of magnitudes were measured from 30 images using *AIP4WIN*. The standard deviations or sigmas of the instrumental V magnitudes were compared with the *AIP4WIN*-derived sigmas, and an equation for the relationship derived, namely:

Actual error = 1.13 * *AIP4WIN* error + 0.007    [3]

It should be noted that Equation 3 will be different for different combinations of telescopes and CCD cameras and observers might wish to perform their own calculations.

For the example here, the total SNR contribution to the error for the three reference stars and the asteroid (see Table 2) is given by: $\sqrt{(0.01^2 + 0.01^2 + 0.02^2 + 0.02^2)} = \pm 0.03$ mag.

#### 3. Combined error

Assuming the component errors are independent, the overall uncertainty (1-sigma standard deviation) is the square root of the sum of the squares of the total SNR error and the reference catalogue error, viz.:

Combined error = $\sqrt{(0.03^2 + 0.03^2)} = \pm 0.04$ mag.

## Colour transformation

If we wish to combine measurements with those of other observers, it is often necessary to measure the small differences between our own telescope/CCD/filter system and the standard filter passbands, in order to correct or transform measurements to a standard magnitude system (in this case Johnson V). The transformation coefficient for RD's setup was ascertained with the help of the BAA Variable Star Section's publication, *Measuring Variable Stars using a CCD camera – a Beginner's Guide*.[22] This method, described in Appendix 6 of that document, uses *Hipparcos* red-blue pairs which are close enough to appear on the same CCD image.

The colour correction is that value which would have to be added to the calculated V magnitude to convert it to a standard V magnitude. Corrections were calculated using the (B–V) values shown in parentheses typical of C-type asteroids (0.7), S-type (0.9), and blue (0.2), red (1.0) and G-type stars (0.6). We ignore extremely red and blue stars in this exercise and where possible use an ensemble of stars to derive the magnitude of the asteroid, so that colour corrections tend to balance out. For RD's system, the average correction that would have to be applied proved to be less than 0.01 mag and can therefore be disregarded.

## Example 2: Determining the V magnitude of asteroid *2000 BD19*

### CMC14-derived V magnitudes compared with absolute photometry based on Hipparcos reference stars

Imaging, calibrations and magnitude calculations described in this example were carried out by RM. As a real-life example, the images have been selected from a campaign to observe the unusual asteroid *2000 BD19* between the dates 2006 Jan 24–Feb 10. This object is unusual in that it has the smallest perihelion distance (0.092 AU) of any object in the solar system for which orbits are accurately known, whilst its aphelion distance is virtually the same as that of Mars.

The observing methodology involved imaging the asteroid in a large telescope (FOV = 8.5×11 arcmin) with no filter, and imaging the same comparison stars together with several stars from the *Hipparcos* catalogue using two wide-field telescopes (FOV = 63×86 arcmin), one equipped with a V filter, the other a Cousins Ic filter. A type of absolute photometry was carried out on the wide-field images, each frame being calibrated in terms of the mean zeropoint, (v–V) of the *Hipparcos* stars adjusted to zero V–Ic colour index using previously derived transformation coefficients for each filter passband. Knowing the V–Ic colour of the comparison stars, it was possible to calculate the V magnitude of these stars to an accuracy of about ±0.015 mag. It is then possible to compare these directly-measured values with CMC14-derived V magnitudes.

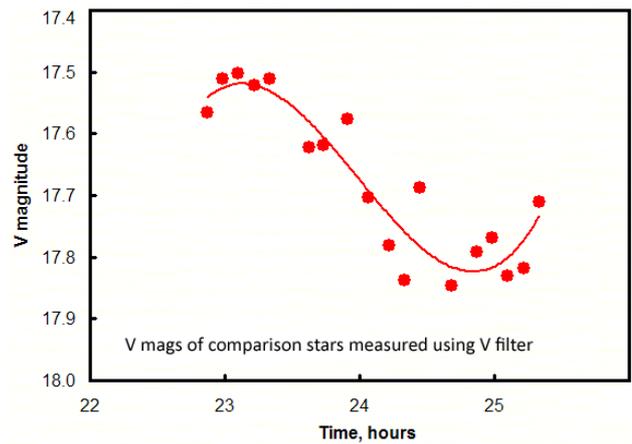

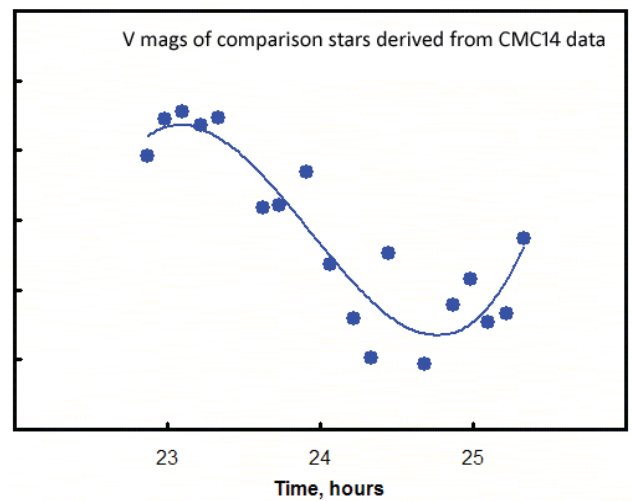

**Figure 4.** An asteroid lightcurve using standard V photometry or using CMC14-derived magnitudes.

### Imaging

For this example, images made on two nights were used. A series of wide-field images was made through filters, namely a stack of 20 90sec V- and Ic-filter exposures on 2006 Jan 25 00:01–00:33UT, and a stack of about 80 60sec V- and Ic-filter exposures on 2006 Jan 29/30 23:42–01:31UT. In the former case, the field was centered near RA 11h 35m, Dec +35.5º, and four *Hipparcos* stars were used as standards (HIP 56516, 56568, 56671 and 56799). In the latter case, the field centre was near RA 11h 18.7m, Dec +41º and four other Hipparcos stars were used as standards (HIP 55101, 55182, 55194 and 55503). A number of suitable comparison stars were identified and their V magnitude determined, values of which are listed in the second column of Tables 4(a) and 4(b). The corresponding CMC14 data were obtained from the catalogue and Equation [2] was employed to derive the V magnitude of each of the stars (14 in all).

It can be seen from Table 4 that the errors (St.dev.) involved using CMC14 data and Equation [2] are on average certainly less than 0.05 mag for a single star, confirming the predicted accuracy from the empirical correlation based on Landolt stars discussed earlier. If 6 or 7 stars are used in an ensemble, then a V magnitude accuracy of ±0.03 mag is possible.

Figure 4 shows two illustrations of the lightcurve of the 17th magnitude asteroid *2000 BD19* depending on whether

**Table 4.** Comparison between measured and calculated V magnitude using Equation [2].

*(a) Comparison stars in the field of asteroid 2000 BD19 on 2006 January 24/25*

| Star ref. | V mag | r' mag | J mag | K mag | J–K colour | $V_{calc}$ | $V-V_{calc}$ |
|---|---|---|---|---|---|---|---|
| a | 13.78 | 13.46 | 12.39 | 11.96 | 0.43 | 13.74 | 0.04 |
| b | 13.28 | 13.06 | 12.16 | 11.83 | 0.33 | 13.27 | 0.01 |
| c | 13.62 | 13.21 | 11.32 | 10.56 | 0.76 | 13.70 | −0.08 |
| d | 13.65 | 13.37 | 12.00 | 11.46 | 0.54 | 13.72 | −0.07 |
| e | 13.88 | 13.66 | 12.62 | 12.23 | 0.39 | 13.91 | −0.03 |
| f | 13.19 | 12.82 | 11.34 | 10.70 | 0.64 | 13.23 | −0.04 |
| g | 13.05 | 12.70 | 11.18 | 10.54 | 0.64 | 13.11 | −0.06 |
| | | | | | | Mean = | −0.031 |
| | | | | | | St.dev. = | 0.044 |

*(b) Comparison stars in the field of asteroid 2000 BD19 on 2006 January 29/30*

| Star ref. | V mag | r' mag | J mag | K mag | J–K colour | $V_{calc}$ | $V-V_{calc}$ |
|---|---|---|---|---|---|---|---|
| d | 10.86 | 10.66 | 9.80 | 9.49 | 0.31 | 10.86 | 0.00 |
| g | 12.67 | 12.45 | 11.68 | 11.41 | 0.27 | 12.62 | 0.05 |
| i | 10.82 | 10.60 | 9.60 | 9.24 | 0.36 | 10.83 | -0.01 |
| j | 11.85 | 11.58 | 10.60 | 10.25 | 0.35 | 11.80 | 0.05 |
| s | 12.70 | 12.45 | 11.41 | 11.03 | 0.38 | 12.69 | 0.01 |
| t | 12.91 | 12.71 | 11.77 | 11.47 | 0.30 | 12.90 | 0.01 |
| w | 12.38 | 11.86 | 10.25 | 9.55 | 0.70 | 12.31 | 0.07 |
| | | | | | | Mean = | 0.024 |
| | | | | | | St.dev. = | 0.030 |

the V magnitudes of comparison stars are measured directly or are calculated using data from the CMC14 catalogue. Each datapoint corresponds to a stack of 10 40sec unfiltered exposures using a 0.28m aperture Schmidt–Cassegrain telescope. Note that although the images were made unfiltered, the SXV-H9 camera has its maximum response close to the V passband, such that if comparison stars are used which are similar in colour to the asteroid, the resultant differential magnitude can be transformed to the V passband. The two lightcurves depicted in Figure 4 are virtually identical, confirming the methodology described here, and showing that the object declined in brightness from maximum to minimum in about 2 hours. Subsequent (unpublished) photometry when the object was brighter showed it to have a rotational period of about 10 hours.

*Finally, a few words of caution!*

Recent work has shown that observations near the Milky Way can lead to misleading results if stars suffer from a significant degree of interstellar reddening by intervening clouds of gas and dust. Imaging asteroids which lie close to the galactic plane is in any case not advisable since the field of view is often extremely crowded, and it is difficult to obtain a measure of the background sky without it being contaminated by faint field stars. Similarly, as it moves the asteroid is continually encountering different field stars which also then contaminate the measuring aperture. As a general rule of thumb, avoid photometry of asteroids if they lie within a galactic latitude of +10 to −10°.

The methodology can be exported to variable star work but in this case it would be helpful to use a large FOV to include as many CMC stars as possible, so that the ensemble value is then largely unaffected by possible intrinsic variability of a few of these stars. Stars fainter than about V=15 should also be avoided to maximise photometric accuracy.

## Conclusion

Equation [2], rounded to an adequate 3 significant figures, can be used to calculate V magnitudes from CMC14 data using the relationship:

$$V = 0.628*(J-K) + 0.995*r'$$

The accuracy of this correlation permits stars down to a V magnitude of 14 to be used to calculate the brightness of an asteroid to within a typical accuracy of ±0.05 mag for a single star. If an ensemble of four or more stars is used then a similar accuracy should be possible down to about V=15. Stars located within ±10° of the galactic equator should be avoided where possible. This approach is a very significant improvement over the use of *Tycho-2* data which has very limited application for asteroid photometry since *Tycho* can only be used down to about V=10.5 with reasonable accuracy.

The advantage of this method is that a good number of comparison stars should be available on even a small CCD image (*e.g.* 12×8 arcmin) thus making the measurement of actual V magnitudes as simple as performing differential photometry. It has particular relevance to the determination of asteroid magnitudes, and hence absolute magnitudes, where the choice of comparison stars varies from night to night as the asteroid tracks across the sky, and the proximity to stars having well-defined magnitudes, *e.g.* Landolt or *Hipparcos*, is far from guaranteed.

## Acknowledgments

The authors are grateful for the assistance given by Gérard Fauré and John Greaves in the preparation of this paper: John for his analysis of *LONEOS* and *2MASS* data and drawing our attention to the formula for the calculation of V magnitude, and Gérard for his encouragement in developing the methodology and suggesting its applicability to measurement of the absolute magnitudes of asteroids. We are also indebted to the referees of the paper, David Boyd and Nick James, for their helpful and constructive comments.

**Addresses:** **RD:** 67 Haslar Crescent, Waterlooville, Hampshire PO7 6DD. [roger.dymock@ntlworld.com]
**RM:** Grange Cottage, Golden Hill, Stourton Caundle, Dorset DT10 2JP. [rmiles@baa.u-net.com]

*Postscript*

# CMC14-based photometry upgrade to the *Astrometrica* software

In the above paper, it has been demonstrated using a few examples that the proposed empirical relationship based on Sloan-r', J and K magnitudes obtained from the CMC14 catalogue will in principle yield an accurate estimate of the V magnitude of field stars in the range, 10<V<15. The relationship is especially valid for stars which are similar in colour to asteroids, so this provides a route whereby accurate V photometry of asteroids is made possible. However, for it to be a practical proposition, some degree of automation is required. Using the current version of Bill Gray's excellent planetarium program, *GUIDE 8.0*, CMC14 star data can be automatically downloaded via the Web enabling suitable comparison stars for any particular asteroid to be readily identified. A simple spreadsheet can then be used to calculate the V magnitudes of these stars, which with suitable photometry software enables asteroid lightcurves to be determined.

In 2008 June, following completion of the draft paper, I received a communication from Herbert Raab, the author of *Astrometrica*, saying that he was working on adding the CMC14 catalogue and wondering if this would be useful for photometry. *Astrometrica* (**http://www.astrometrica.at/**) is a very popular piece of software used to carry out astrometry of asteroids and comets, but the photometry it yields when based on catalogues such as the USNO B1.0 and UCAC-2 is only accurate to about 0.2 mag at best. To cut a long story short, some 10 weeks and 11 revisions of the software later, *Astrometrica* was rewritten with some help from me to permit photometry accurate to about ±0.03 mag in those regions of the sky covered by CMC14. The latest version (4.5.1.377) provides for aperture photometry in both the Johnson-V and Cousins-R passbands. The relationship described in the present paper is utilised by *Astrometrica* along with constraints on the J–K colours of reference stars, to yield V photometry to a precision of 0.01 magnitude. Note that the CMC14 catalogue option must be selected in the Settings configuration file for accurate photometry. Cousins-R photometry is based on the relationship, R = r'–0.22.

In 2008 September, the binary asteroid 2000 $DP_{107}$ made a close approach to the Earth and I was able to secure a series of about 500 30sec exposures on the night of September 26/27 from Golden Hill Observatory, Dorset (MPC Code J77). This activity is in support of the Ondrejov Observatory Survey of Binary Asteroids, which is led by Dr Petr Pravec.

During the observing run, the 15th magnitude object traversed three different fields of view. The image frames were processed using *Astrometrica*, which both dark-subtracts and flat-fields each frame in turn before identifying an ensemble of suitable stars, which are then used *en masse* to determine the V magnitude and exact position of the moving asteroid. The resultant lightcurve comprising 462 datapoints is depicted in Figure A1.

It can be seen that a very acceptable result has been obtained given the faintness of the object and the size of the telescope used (28cm aperture). The irregular nature of the lightcurve is a consequence of this being a binary system, comprising a primary body some 800 metres across which rotates every 2.78 hours, around which orbits a secondary object about 350m in size, completing a single revolution every 42.2 hours. The secondary is thought to be locked in ▶ synchronous rotation about the primary (rather like the Earth–Moon system) and occasionally undergoes mutual eclipses and occultations. One such event, lasting about 2 hours, happened to take place during my observing run of September 26/27: the start and end of the event have been highlighted in Figure A1. Since the form of the lightcurve outside of eclipse/occultation is relatively complex, this has been subtracted by Petr Pravec from the raw lightcurve to produce Figure A2, which depicts the contribution arising from the mutual event alone.

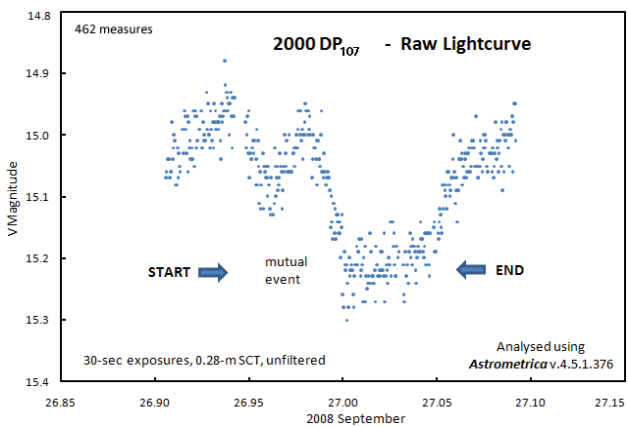

**Figure A1.** Raw lightcurve of the binary asteroid 2000 $DP_{107}$, produced using *Astrometrica* upgraded for CMC14 photometry.

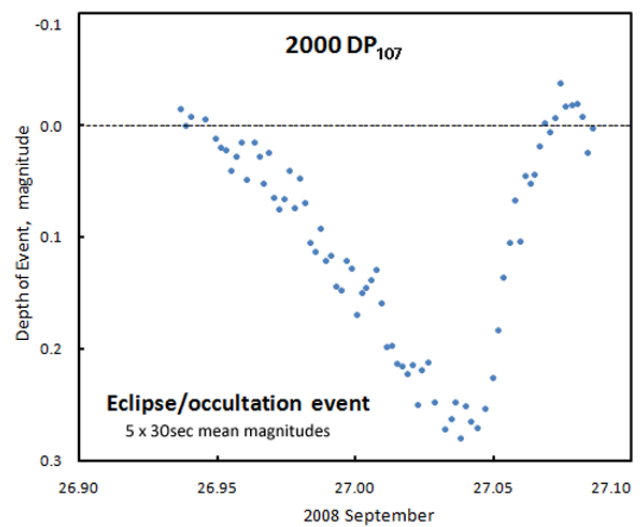

**Figure A2.** Lightcurve showing details of an eclipse/occultation event, obtained after subtracting the normal rotational lightcurve of the binary system from the data shown in Figure A1.

Given that the target was 15th magnitude at the time, the form of the event is very well-defined and clearly demonstrates success in upgrading *Astrometrica* for photometry of asteroids. It should be noted that in the majority of cases, the use of two different configuration file settings in *Astrometrica* is preferred: one optimised for astrometry and one for photometry. This is especially true if the region of sky traversed by the asteroid is very crowded with field stars.

I would like to express my sincerest gratitude to Herbert Raab for his patience and perseverance in making the necessary modifications to his software – I am most grateful to him.

**Richard Miles** *[rmiles@baa.u-net.com]*
*Director, Asteroids and Remote Planets Section*

*2008 November 14*